\documentclass[preprint,12pt]{elsarticle}

\usepackage{amssymb}
\usepackage{amsmath,amssymb,amscd,amsbsy,amsgen,amsopn,amstext,amsxtra}
\usepackage[mathscr]{eucal}

\newcommand{\vc}{\boldsymbol}
\newcommand{\s}{\:\!}

\begin{document}

\begin{frontmatter}



\title{Relativistic Lagrangians for the Lorentz-Dirac equation}


\author[Deg]{Shinichi Deguchi\corref{mycorrespondingauthor}}
\cortext[mycorrespondingauthor]{Corresponding author.}
\ead{deguchi@phys.cst.nihon-u.ac.jp}

\address[Deg]{Institute of Quantum Science, College of Science and Technology, 
Nihon University, Chiyoda-ku, Tokyo 101-8308, Japan}

\author{Kunihiko Nakano${}^{\rm a}$}

\author[Suz]{Takafumi Suzuki}

\address[Suz]{Junior College Funabashi Campus, 
Nihon University, Narashinodai, Funabashi, Chiba 274-8501, Japan}

\begin{abstract}
We present two types of relativistic Lagrangians for the Lorentz-Dirac equation 
written in terms of an arbitrary world-line parameter.  
One of the Lagrangians contains an exponential damping function of the proper time 
and explicitly depends on the world-line parameter.  
Another Lagrangian includes additional cross-terms consisting of auxiliary dynamical variables 
and does not depend explicitly on the world-line parameter. 
We demonstrate that both the Lagrangians actually yield the Lorentz-Dirac equation with a source-like term. 
\end{abstract}

\end{frontmatter}

\numberwithin{equation}{section}

\section{Introduction}

A charged particle emitting electromagnetic radiation 
is subjected to the reaction force caused by the  
particle's own electromagnetic radiation. This phenomenon is well-known as the {\em radiation reaction} 
\cite{LL, Jackson, Barut1, Rohrlich1, Rohrlich2, Spohn1}. 
It was first evaluated by Lorentz at the end of the 19th century \cite{Lorentz1} 
and subsequently argued by Abraham and Lorentz  
on the basis of the charged rigid sphere model of a charged particle \cite{Abraham, Lorentz2}. 
In the zero radius limit that this model tends to become a point charge, 
the classical non-relativistic equation of motion 
for the charged particle located at the position $\vc{x}=\vc{x}(t)$ is found to be 
\begin{align}
m\frac{d^2 \vc{x}}{dt^2}=\vc{F} +\frac{2}{3} e^{2} \frac{d^3 \vc{x}}{dt^3} \,, 
\label{1}
\end{align}
where $m$ is the physical mass of the particle, $e$ its electric charge, 
and $\vc{F}$ denotes the external Lorentz force.  
(In this paper, we employ units such that $c=1$.) 
Equation (\ref{1}) is called the Lorentz-Abraham equation 
(or the Abraham-Lorentz equation). 
A relativistic extension of the Lorentz-Abraham equation was derived by Dirac 
in a manifestly covariant manner by considering energy-momentum conservation \cite{Dirac},       
and is now often called the Lorentz-Dirac equation 
\cite{Barut1, Spohn1, Barut2, Poisson, Nakhleh}.  
With the spacetime coordinates $x^{\mu}=x^{\mu}(l)$ ($\mu=0,1,2,3$) of a charged  
particle propagating in 4-dimensional Minkowski space, the Lorentz-Dirac equation reads 
\begin{align}
m \frac{du^{\mu}}{dl} =eF^{\mu\nu}(x) {u}_{\nu} 
+\frac{2}{3} e^{2}\big( \delta^{\mu}{}_{\nu} -u^{\mu} u_{\nu} \big) \frac{d^2 u^{\nu}}{dl^{2}} \,. 
\label{2}
\end{align}
Here, $u^{\mu}:=dx^{\mu}/dl$, 
$F^{\mu\nu}$ is the field strength tensor of an external electromagnetic field, 
and $l$ denotes the proper time of the particle or, in other words,  
the arc length of the world-line traced out by the particle.    
The metric tensor of Minkowski space is assumed to be $\eta_{\mu\nu}=\mathrm{diag}(1,-1,-1,-1)$, 
so that $u_{\mu} u^{\mu}=\eta_{\mu\nu} u^{\mu} u^{\nu}=1$ holds.  
Equations (\ref{1}) and (\ref{2}) are unusual ones including third-order time derivatives of 
the particle's position coordinates. 
In connection with this fact, these equations admit physically unacceptable solutions 
such as runaway and pre-acceleration solutions 
\cite{LL, Jackson, Rohrlich1, Rohrlich2,  Spohn1, Poisson, Nakhleh, Yaghjian}. 
To overcome this problem, various ideas have been proposed until recently 
\cite{LL, Spohn1, Nakhleh, Yaghjian, Barut3, FOC1, FOC2, Spohn2, Rohrlich3, Rohrlich4, Seto}; 
however, it seems that an ultimate solution to the problem has not been found yet.

Once the equations of motion (\ref{1}) and (\ref{2}) have been obtained, it is quite natural to seek 
Lagrangians corresponding to these equations in order to develop the Lagrangian and Hamiltonian 
formulations of a charged particle subjected to the radiation reaction force. 
If these formulations are established, they might lead to a novel quantum-mechanical description of 
a charged particle undergoing radiation reaction and might give us new room to deal with 
the above-mentioned problem. 
As far as the present authors know, 
there have been a few attempts to construct Lagrangians 
corresponding to Eqs. (\ref{1}) and (\ref{2}) until now \cite{Carati, BM, Kupriyanov}. 
Carati constructed an explicitly time-dependent Lagrangian for Eq (\ref{1}) 
with the use of auxiliary dynamical variables \cite{Carati}. 
(Carati also considered a relativistic extension of this Lagrangian in an extremely limited case.) 
Barone and Mendes derived an explicitly time-independent Lagrangian for Eq. (\ref{1}) by incorporating  
the time-reversed copy of Eq. (\ref{1}) into the original setting \cite{BM}. 
Carati's and Barone-Mendes's approaches are, respectively, based on learning from
the direct and indirect Lagrangian formulations of the damped harmonic oscillator 
\cite{Dekker, Caldirola, Kanai, Bateman, FT, Banerjee}.\footnote{~
The direct formulation adopts an explicitly time-dependent Lagrangian of 
the damped harmonic oscillator \cite{Dekker, Caldirola, Kanai}, 
while the indirect formulation adopts an explicitly time-independent Lagrangian for a system 
consisting of the damped harmonic oscillator and its time-reversed counterpart 
\cite{Dekker, Bateman, FT, Banerjee}.}   
It should be pointed out here that in these approaches, the external Lorentz force 
$\vc{F}$ is assumed to be independent of the velocity $\vc{v}:=d\vc{x}/dt$. 
Hence, it follows that in actuality, Carati's and Barone-Mendes's Lagrangians 
can describe only a charged particle being in the particular situation in which the magnetic field vanishes 
or is parallel to $\vc{v}$.\footnote{~
In Ref. \cite{Kupriyanov}, Kupriyanov investigated the possibility of constructing Lagrangians  
corresponding to Eqs. (\ref{1}) and (\ref{2}) and reached the conclusion that 
there exist no corresponding Lagrangians. 
However, Kupriyanov's proof of this conclusion considers Lagrangians consisting only of 
the coordinate variables, such as $\vc{x}$ and $x^{\mu}$, and their first- and second-order time derivatives.   
Since Carati's and Barone-Mendes's Lagrangians contain extra dynamical variables, 
these Lagrangians are outside the scope of Kupriyanov's proof.}

In this paper, we present two types of Lagrangians for the Lorentz-Dirac equation (\ref{2}) 
that are constructed in such a fashion that the corresponding actions 
remain invariant under reparametrization of a world-line parameter along the particle's world-line.   
These Lagrangians are completely relativistic and admit the general form of the external Lorentz force.  
Also, the Lagrangians are outside the scope of Kupriyanov's proof \cite{Kupriyanov}, because they 
contain auxiliary dynamical variables in addition to $x^{\mu}$. 
One of the Lagrangians contains an exponential damping function of the proper time $l$, 
while another Lagrangian includes additional cross-terms consisting of two auxiliary dynamical variables.  
Both the Lagrangians include terms similar to what can be  
seen in the Lagrangian that governs a certain model of a relativistic point particle with rigidity 
\cite{Pav1, Pav2, Plyushchay}.  
We would like to emphasize that our Lagrangians are not immediate extensions of 
Carati's and Barone-Mendes's Lagrangians.

This paper is organized as follows. In section 2, we introduce necessary dynamical variables 
and define their transformation rules under reparametrization of a world-line parameter. 
In section 3, we present a Lagrangian that contains an exponential damping function  
and show that the Lagrangian actually yields the Lorentz-Dirac equation with a source-like term. 
In section 4, we consider a Lagrangian including additional cross-terms, 
instead of the exponential damping function, 
and show that this Lagrangian also yields the Lorentz-Dirac equation with a source-like term. 
Section 5 is devoted to a summary and discussion. 
Appendix A provides the Lorentz-Dirac equation written in terms of an arbitrary world-line parameter 
instead of the proper time $l$.

\section{Preliminaries: dynamical variables and their transformation rules}

Let $\tau$ $(\tau_{0}\le\tau\le\tau_{1})$ be an arbitrary world-line parameter along the particle's world-line, 
being chosen in such a manner that $dx^{0}/d\tau >0$. 
The spacetime coordinates of a charged particle are now denoted as $x^{\mu}=x^{\mu}(\tau)$. 
Under the reparametrization $\tau \rightarrow \tau^{\prime}=\tau^{\prime}(\tau)$ ($d\tau^{\prime}/d\tau >0$),  
the coordinate variables $x^{\mu}$ behave as scalar fields on the 1-dimensional parameter space  
$\mathcal{T}:=\{  \tau\;\! |\, \tau_{0} \leq \tau \leq \tau_{1} \}\:\!$:   
\begin{align}
x^{\mu}(\tau) \rightarrow x^{\prime\mu}(\tau^{\prime})=x^{\mu}(\tau) \,. 
\label{2.1}
\end{align}
In addition to $x^{\mu}$, we introduce auxiliary dynamical variables 
$q_{i}^{\mu}=q_{i}^{\mu}(\tau)$, $\lambda_{i\mu}=\lambda_{i\mu}(\tau)$ $(i=1,2)$, 
and $\xi_{\mu}=\xi_{\mu}(\tau)$.  
They are assumed to transform under the reparametrization as 
scalar-density fields of weight 1 on $\mathcal{T}\:\!$: 
\begin{align}
q^{\mu}_{i}(\tau)& \rightarrow q^{\prime\mu}_{i} (\tau^{\prime})
=\frac{d\tau}{d\tau^{\prime}}q^{\mu}_{i} (\tau) \,,
\label{2.2}
\\
\lambda_{i\mu}(\tau) & \rightarrow \lambda^{\prime}_{i\mu}(\tau^{\prime})
=\frac{d\tau}{d\tau^{\prime}}  \lambda_{i\mu}(\tau) \,, 
\label{2.3}
\\
\xi_{\mu}(\tau) & \rightarrow \xi^{\prime}_{\mu}(\tau^{\prime})
=\frac{d\tau}{d\tau^{\prime}}  \xi_{\mu}(\tau) \,.
\label{2.4}
\end{align}
The components of the vector resolute of the 4-vector $(\dot{q}_{i}^{\mu})$ 
perpendicular to $(q_{i}^{\mu})$ are given by 
\begin{align}
\dot{q}_{i\perp}^{\mu} :=\dot{q}_{i}^{\mu}-\frac{q_{i} \dot{q}_{i}}{q_{i}^{2}} {q}_{i}^{\mu} ,
\label{2.5}
\end{align}
where $\dot{q}_{i}^{\mu}:=d{q}_{i}^{\mu}/d\tau$, $q_{i}^{2}:=q_{i\mu} q_{i}^{\mu}$, and 
$q_{i} \dot{q}_{i}:=q_{i\mu} \dot{q}_{i}^{\mu}$ (no sum with respect to $i^{\;\!}$). 
It can be shown by using Eq. (\ref{2.2}) that unlike $\dot{q}_{i}^{\mu}$, 
the components $\dot{q}_{i\perp}^{\mu}$ transform homogeneously as 
\begin{align}
\dot{q}_{i\perp}^{\mu}(\tau) & \rightarrow \dot{q}_{i\perp}^{\prime\mu}(\tau^{\prime})
=\left(\frac{d\tau}{d\tau^{\prime}} \right)^{2} \dot{q}_{i\perp}^{\mu}(\tau) \,. 
\label{2.6}
\end{align}
We thus see that under the reparametrization, $\dot{q}_{i\perp}^{\mu}$ behave 
as scalar-density fields of weight 2 on $\mathcal{T}$.

\section{A Lagrangian with an exponential damping function} 

Now, from the dynamical variables $x^{\mu}$, $q_{i}^{\mu}$, $\lambda_{i\mu}$, and $\xi_{\mu}$, 
we construct the following Lagrangian:  
\begin{align}
L_{\rm D} &=\frac{\exp(-kl \s)}{\left( q_{1}^{2} q_{2}^{2} \right)^{1/4}} 
\left[\;\! \frac{1}{2} \left( \frac{\dot{q}_{1\perp}^{2}}{q_{1}^{2}} -\frac{\dot{q}_{2\perp}^{2}}{q_{2}^{2}} \right) \right. 
\notag
\\
& \quad\, 
\left. 
-\lambda_{1\mu} \left( q_{1}^{\mu}-\dot{x}^{\mu} \right) 
+\lambda_{2\mu} \left(q_{2}^{\mu}-\dot{x}^{\mu} \right) 
+\xi_{\mu} \left( q_{1}^{\mu} -q_{2}^{\mu} \right) 
-\frac{3}{2e} F_{\mu\nu}(x) q_{1}^{\mu} q_{2}^{\nu} \:\! \right] , 
\label{3.1}
\end{align}
where $k:=3m/2e^{2}$, $\dot{x}^{\mu}:=dx^{\mu}/d\tau$, 
$\dot{q}_{i\perp}^{2}:=\dot{q}_{i\perp\mu} \dot{q}^{\mu}_{i\perp}$, 
and $F_{\mu\nu}(=-F_{\nu\mu})$ is again the field strength tensor of an external electromagnetic field. 
In Eq. (\ref{3.1}), the proper time $l$ is a function of $\tau$ represented as 
\begin{align}
l(\tau)=\int_{\tau_{0}}^{\tau} d\tilde{\tau} 
\sqrt{\dot{x}_{\mu}(\tilde{\tau}) \dot{x}^{\mu}(\tilde{\tau})} \:. 
\label{3.2}
\end{align} 
Here, $x^{\mu}(\tilde{\tau})$ is understood as a solution of 
the equation of motion for $x^{\mu}$ obtained later,  
not as a dynamical variable whose variation is taken into account in varying the action 
\begin{align}
S_{\rm D}=\int_{\tau_0}^{\tau_1} d\tau L_{\rm D} \,. 
\label{3.3}
\end{align}
The Lagrangian $L_{\rm D}$ explicitly depends on $\tau$ via  
the exponential damping function $\exp(-kl^{\s})$. 
Since $l(\tau)$ is geometrically the arc length of the particle's world-line, 
it is certainly reparametrization invariant.\footnote{~Strictly speaking, 
$l(\tau)$ is a functional of $x^{\mu}$ as well as a function of $\tau$ and $\tau_0$. 
In this sense, $l(\tau)$ should be read as $l(\tau, \tau_0\:\!; x^{\mu})$.   
The reparametrization invariance of $l(\tau)$ can be expressed as 
$l(\tau^{\prime}, \tau_{0}^{\prime}\:\! ; x^{\prime \mu})=l(\tau, \tau_0\:\!; x^{\mu})$.} 
Considering this fact and using the transformation rules in 
Eqs. (\ref{2.1}), (\ref{2.2}), (\ref{2.3}), (\ref{2.4}), and (\ref{2.6}), 
we can show that the action $S_{\rm D}$ remains invariant under the reparametrization  
$\tau \rightarrow \tau^{\prime}$. 
We also see that $L_{\rm D}$ remains invariant under the gauge transformation 
\begin{align} 
\lambda_{1\mu} \rightarrow \lambda_{1\mu}^{\prime}=\lambda_{1\mu}+\theta_{\mu} \,, 
\quad 
\lambda_{2\mu} \rightarrow \lambda_{2\mu}^{\prime}=\lambda_{2\mu}+\theta_{\mu} \,, 
\quad
\xi_{\mu} \rightarrow \xi_{\mu}^{\prime}=\xi_{\mu}+\theta_{\mu} \,, 
\label{3.4}
\end{align}
with real gauge functions $\theta^{\mu} =\theta^{\mu}(\tau)$.  
The Lagrangian $L_{\rm D}$ has the antisymmetric property 
\begin{align}
L_{\rm D} \left(q_{1}^{\mu}, \dot{q}_{1}^{\mu}, \lambda_{1\mu}; 
q_{2}^{\mu}, \dot{q}_{2}^{\mu}, \lambda_{2\mu} \right)
=-L_{\rm D} \left(q_{2}^{\mu}, \dot{q}_{2}^{\mu}, \lambda_{2\mu}; 
q_{1}^{\mu}, \dot{q}_{1}^{\mu}, \lambda_{1\mu} \right) . 
\label{3.5}
\end{align}

Let us derive the Euler-Lagrange equations for the dynamical variables from $L_{\rm D}$.  
Noting that $x^{\mu}(\tilde{\tau})$ contained in $l(\tau)$ and hence $l(\tau)$ itself     
are not objects for taking variation, 
we can easily obtain the Euler-Lagrange equation for $x^{\mu}\:\!$: 
\begin{align}
\frac{d}{d\tau} \left[ \frac{\exp(-kl \s)}{\left( q_{1}^{2} q_{2}^{2} \right)^{1/4}}
\left(\lambda_{1\mu}-\lambda_{2\mu} \right) \right] 
+\frac{3\exp(-kl \s)}{2e \left( q_{1}^{2} q_{2}^{2} \right)^{1/4}} 
\;\! \partial_{\mu} F_{\nu\rho}(x) q_{1}^{\nu} q_{2}^{\rho} =0\,.
\label{3.6}
\end{align}
This equation includes the gauge-invariant quantity $\lambda_{1\mu}-\lambda_{2\mu}$ 
as a reflection of the gauge invariance of $L_{\rm D}$. 
Hence, $\lambda_{1\mu}$ and $\lambda_{2\mu}$ themselves are not uniquely determined. 
The Euler-Lagrange equation for $q_{1}^{\mu}$ can be written as 
\begin{align}
&\frac{\exp(-kl \s)}{\left( q_{1}^{2} q_{2}^{2} \right)^{1/4}} 
\left[\;\! \frac{1}{2} \left( \frac{d}{d\tau} \frac{\partial K_{1}}{\partial\dot{q}_{1}^{\mu}}
-\frac{\partial K_{1}}{\partial q_{1}^{\mu}} \right)
+\lambda_{1\mu} -\xi_{\mu} +\frac{3}{2e} F_{\mu\nu}(x) q_{2}^{\nu} \:\! \right] 
\notag
\\
&
+\left(\frac{d}{d\tau} \frac{\exp(-kl \s)}{\left(q_{1}^{2} q_{2}^{2} \right)^{1/4}} \right) 
\frac{1}{2} \frac{\partial K_{1}}{\partial\dot{q}_{1}^{\mu}} 
+\frac{q_{1\mu}}{2q_{1}^{2}} L_{\rm D} =0 \,,  
\label{3.7}
\end{align}
with
\begin{align}
K_{1}:=\frac{\dot{q}_{1\perp}^{2}}{q_{1}^{2}}
=\frac{q_{1}^{2} \dot{q}_{1}^{2} -\left( q_{1} \dot{q}_{1} \right)^{2}}{\left(q_{1}^{2} \right)^{2}}\,.
\label{3.8}
\end{align}
Applying the formulas 
\begin{align}
&\frac{1}{2} \frac{\partial K_{1}}{\partial\dot{q}_{1}^{\mu}}
=\frac{\dot{q}_{1\perp\mu}}{q_1^{2}} 
=\frac{1}{\sqrt{q_{1}^{2}}}\:\! \frac{d}{d\tau} \frac{q_{1\mu}}{\sqrt{q_{1}^{2}}} \, , 
\label{3.9}
\\
&\frac{d}{d\tau} \frac{\partial K_{1}}{\partial\dot{q}_{1}^{\mu}}
-\frac{\partial K_{1}}{\partial q_{1}^{\mu}} 
=\frac{2}{q_{1}^{2}} \left( \:\!\ddot{q}_{1\perp\mu}-\frac{2q_{1} \dot{q}_{1}}{q_{1}^{2}} \dot{q}_{1\perp\mu} \right) , 
\label{3.10}
\\
&\frac{d}{d\tau} \frac{1}{\left( q_{1}^{2} q_{2}^{2} \right)^{1/4}}
=-\frac{1}{2\left( q_{1}^{2} q_{2}^{2} \right)^{1/4}} 
\left( \frac{q_{1} \dot{q}_{1}}{q_{1}^{2}} +\frac{q_{2} \dot{q}_{2}}{q_{2}^{2}} \right) 
\label{3.11}
\end{align}
to Eq. (\ref{3.7}) appropriately, we obtain 
\begin{align}
k\frac{dl}{d\tau} 
\frac{1}{\sqrt{q_{1}^{2}}}\:\! \frac{d}{d\tau} \frac{q_{1\mu}}{\sqrt{q_{1}^{2}}} 
&=\frac{3}{2e} F_{\mu\nu}(x) q_{2}^{\nu} 
+\frac{\left( q_{1}^{2} q_{2}^{2} \right)^{1/4} q_{1\mu}}{2q_1^2 \exp(-kl \s)} L_{\rm D}
\notag
\\
& \,\quad
+\frac{\ddot{q}_{1\perp\mu}}{q_{1}^{2}} 
-\left( \frac{5q_{1} \dot{q}_{1}}{q_{1}^{2}} +\frac{q_{2} \dot{q}_{2}}{q_{2}^{2}} \right) 
\frac{\dot{q}_{1\perp \mu}}{2q_1^2}
+\lambda_{1\mu} -\xi_{\mu} \,. 
\label{3.12}
\end{align}
Here, $\ddot{q}_{1\perp\mu}$, together with $\ddot{q}_{2\perp\mu}$, is defined by 
\begin{align}
\ddot{q}_{i\perp\mu} :=\ddot{q}_{i\mu}-\frac{q_{i} \ddot{q}_{i}}{q_{i}^{2}} {q}_{i\mu} \,,
\label{3.13}
\end{align}
where $\ddot{q}_{i\mu}:=d^2 {q}_{i\mu}/d\tau^2$ and 
$q_{i} \ddot{q}_{i}:=q_{i\mu} \ddot{q}_{i}^{\mu}$ (no sum with respect to $i^{\;\!}$). 
Following the same procedure as that used for deriving Eq. (\ref{3.12}), 
we can derive the Euler-Lagrange equation for $q_{2}^{\mu}$ as 
\begin{align}
k\frac{dl}{d\tau} 
\frac{1}{\sqrt{q_{2}^{2}}}\:\! \frac{d}{d\tau} \frac{q_{2\mu}}{\sqrt{q_{2}^{2}}} 
&=\frac{3}{2e} F_{\mu\nu}(x) q_{1}^{\nu} 
-\frac{\left( q_{1}^{2} q_{2}^{2} \right)^{1/4} q_{2\mu}}{2q_2^2 \exp(-kl \s)} L_{\rm D}
\notag
\\
& \,\quad
+\frac{\ddot{q}_{2\perp\mu}}{q_{2}^{2}} 
-\left( \frac{q_{1} \dot{q}_{1}}{q_{1}^{2}} +\frac{5q_{2} \dot{q}_{2}}{q_{2}^{2}} \right) 
\frac{\dot{q}_{2\perp \mu}}{2q_2^2}
+\lambda_{2\mu} -\xi_{\mu} \,. 
\label{3.14}
\end{align}
The Euler-Lagrange equations for $\lambda_{1\mu}$, $\lambda_{2\mu}$, and $\xi_{\mu}$ 
are respectively found to be  
\begin{align}
q_{1}^{\mu}&=\dot{x}^{\mu} ,
\label{3.15}
\\
q_{2}^{\mu}&=\dot{x}^{\mu} , 
\label{3.16}
\\
q_{1}^{\mu}&=q_{2}^{\mu} \;\! . 
\label{3.17}
\end{align}
Equation (\ref{3.17}) can also be found from Eqs. (\ref{3.15}) and (\ref{3.16}).

Substituting Eqs. (\ref{3.15}) and (\ref{3.16}) into Eq. (\ref{3.12}) and noting 
\begin{align}
L_{\rm D} \left(q_{1}^{\mu}, \dot{q}_{1}^{\mu}, \lambda_{1\mu}; 
q_{2}^{\mu}, \dot{q}_{2}^{\mu}, \lambda_{2\mu} \right)
=L_{\rm D} \left(\dot{x}^{\mu}, \ddot{x}^{\mu}, \lambda_{1\mu}; 
\dot{x}^{\mu}, \ddot{x}^{\mu}, \lambda_{2\mu} \right) 
=0\,, 
\label{3.18}
\end{align}
we have 
\begin{align}
k\frac{dl}{d\tau} 
\frac{1}{\sqrt{\dot{x}^{2}}}\:\! \frac{d}{d\tau} \frac{\dot{x}^{\mu}}{\sqrt{\dot{x}^{2}}} 
&=\frac{3}{2e} F^{\mu\nu}(x) \dot{x}_{\nu}  
+\frac{\dddot{x}_{\perp}^{\mu}}{\dot{x}^{2}} 
-\frac{3 (\dot{x} \ddot{x}) \ddot{x}_{\perp}^{\mu}}{(\dot{x}^{2})^{2}} 
+\lambda_{1}^{\mu} -\xi^{\mu} \,, 
\label{3.19}
\end{align}
where $\ddot{x}^{\mu}:=d^2 x^{\mu}/d\tau^2$ and $\dddot{x}^{\mu}:=d^3 x^{\mu}/d\tau^3$. 
Similarly, substituting Eqs. (\ref{3.15}) and (\ref{3.16}) into Eq. (\ref{3.14}) and using (\ref{3.18}), 
we have 
\begin{align}
k\frac{dl}{d\tau} 
\frac{1}{\sqrt{\dot{x}^{2}}}\:\! \frac{d}{d\tau} \frac{\dot{x}^{\mu}}{\sqrt{\dot{x}^{2}}} 
&=\frac{3}{2e} F^{\mu\nu}(x) \dot{x}_{\nu}  
+\frac{\dddot{x}_{\perp}^{\mu}}{\dot{x}^{2}} 
-\frac{3 (\dot{x} \ddot{x}) \ddot{x}_{\perp}^{\mu}}{(\dot{x}^{2})^{2}} 
+\lambda_{2}^{\mu} -\xi^{\mu} \,. 
\label{3.20}
\end{align}
Comparing Eq. (\ref{3.19}) with Eq. (\ref{3.20}) leads to 
\begin{align}
\lambda_{1}^{\mu}=\lambda_{2}^{\mu} \,. 
\label{3.21}
\end{align}
This equality is covariant under the gauge transformation (\ref{3.4}). 
It follows from Eq. (\ref{3.21}) that Eq. (\ref{3.6}) is identically satisfied, because 
$\partial_{\mu} F_{\nu\rho}(x) q_{1}^{\nu} q_{2}^{\rho}
=\partial_{\mu} F_{\nu\rho}(x) \dot{x}^{\nu} \dot{x}^{\rho}=0$ holds by using Eqs. (\ref{3.15}) and (\ref{3.16}). 
Hereafter, taking into account Eq. (\ref{3.21}), 
we simply write $\lambda_{1}^{\mu}$ and $\lambda_{2}^{\mu}$ as $\lambda^{\mu}$. 
Thereby, Eqs (\ref{3.19}) and (\ref{3.20}) can be written together as a single equation 
\begin{align}
k\frac{dl}{d\tau} 
\frac{1}{\sqrt{\dot{x}^{2}}}\:\! \frac{d}{d\tau} \frac{\dot{x}^{\mu}}{\sqrt{\dot{x}^{2}}} 
&=\frac{3}{2e} F^{\mu\nu}(x) \dot{x}_{\nu}  
+\frac{\dddot{x}_{\perp}^{\mu}}{\dot{x}^{2}} 
-\frac{3 (\dot{x} \ddot{x}) \ddot{x}_{\perp}^{\mu}}{(\dot{x}^{2})^{2}} 
+\varLambda^{\mu} ,  
\label{3.22}
\end{align}
where $\varLambda^{\mu} :=\lambda^{\mu} -\xi^{\mu}$. 
Obviously, $\varLambda^{\mu}$ is invariant under the gauge transformation (\ref{3.4}). 
Equation (\ref{3.22}) is precisely the equation of motion for $x^{\mu}$ mentioned under Eq. (\ref{3.2}). 
Since $x^{\mu}$ contained in $l$ has been assumed to be a solution of Eq. (\ref{3.22}), 
it can be identified with $x^{\mu}$ in Eq. (\ref{3.22}).   
Upon considering this fact, substituting the $\tau$-derivative of Eq. (\ref{3.2}), 
i.e., $dl(\tau)/d\tau =\sqrt{\dot{x}_{\mu}(\tau) \dot{x}^{\mu}(\tau)}^{\:\!}$, into Eq. (\ref{3.22}) 
and recalling $k:=3m/2e^{2}$,  
we obtain 
\begin{align}
m \frac{d}{d\tau} \frac{\dot{x}^{\mu}}{\sqrt{\dot{x}^{2}}} 
&=eF^{\mu\nu}(x) \dot{x}_{\nu}  
+\frac{2}{3}e^{2} \left( \frac{\dddot{x}_{\perp}^{\mu}}{\dot{x}^{2}} 
-\frac{3 (\dot{x} \ddot{x}) \ddot{x}_{\perp}^{\mu}}{(\dot{x}^{2})^{2}} \right) 
+\frac{2}{3}e^{2} \varLambda^{\mu} . 
\label{3.23}
\end{align}
If $\varLambda^{\mu}=0$, Eq. (\ref{3.23}) is identical with 
the Lorentz-Dirac equation written in terms of the arbitrary world-line parameter $\tau^{\:\!}$;   
see Eq. (\ref{A8}) in Appendix A.    
For this reason, Eq. ({\ref{3.23}) can be said to be  
the Lorentz-Dirac equation with a source-{\em like} term $2e^{2} \varLambda^{\mu}/3$. 
We thus see that the Lagrangian $L_{\rm D}$ yields 
the Lorentz-Dirac equation with a source-like term.

Now we adopt the proper-time gauge $\tau=l$, choosing $\tau$ to be the proper time $l$. 
Accordingly, $\dot{x}^{\mu}=u^{\mu}$, $\dot{x}^{2}=1$, and $\dot{x} \ddot{x}=0$ 
are valid, so that Eq. (\ref{3.23}) becomes 
\begin{align}
m \frac{du^{\mu}}{dl} =eF^{\mu\nu}(x) {u}_{\nu} 
+\frac{2}{3} e^{2}\big( \delta^{\mu}{}_{\nu} -u^{\mu} u_{\nu} \big) \frac{d^2 u^{\nu}}{dl^{2}} 
+\frac{2}{3}e^{2} \varLambda^{\mu} .
\label{3.24}
\end{align}
This is exactly what is defined by adding the source-like term  
$2e^{2} \varLambda^{\mu}/3$ to the (original) Lorentz-Dirac equation (\ref{2}).

\section{A Lagrangian with additional cross-terms} 

Next we consider an alternative Lagrangian defined by 
\begin{align}
L_{\rm A} &=\frac{1}{\left( q_{1}^{2} q_{2}^{2} \right)^{1/4}} 
\left[\;\! \frac{1}{2} \left( \frac{\dot{q}_{1\perp}^{2}}{q_{1}^{2}} -\frac{\dot{q}_{2\perp}^{2}}{q_{2}^{2}} \right) 
-\frac{k}{2} \left( \frac{\dot{q}_{1\perp\mu} q_{2}^{\mu}}{\sqrt{q_{1}^{2}}} 
-\frac{\dot{q}_{2\perp\mu} q_{1}^{\mu}}{\sqrt{q_{2}^{2}}} \right) \right. 
\notag 
\\
& \quad\, 
\left. 
-\lambda_{1\mu} \left( q_{1}^{\mu}-\dot{x}^{\mu} \right) 
+\lambda_{2\mu} \left(q_{2}^{\mu}-\dot{x}^{\mu} \right) 
+\xi_{\mu} \left( q_{1}^{\mu} -q_{2}^{\mu} \right) 
-\frac{3}{2e} F_{\mu\nu}(x) q_{1}^{\mu} q_{2}^{\nu} \:\! \right] .
\label{4.1}
\end{align}
Here it should be emphasized that $L_{\rm A}$ includes the additional cross-terms proportional to $k$ 
instead of the exponential damping function $\exp(-kl^{\s})$.  
Also, it is worth noting that unlike $L_{\rm D}$, the Lagrangian $L_{\rm A}$  
does not depend explicitly on $\tau$. 
Using the transformation rules in Eqs. (\ref{2.1}), (\ref{2.2}), (\ref{2.3}), (\ref{2.4}), and (\ref{2.6}), 
we can show that the action 
\begin{align}
S_{\rm A}=\int_{\tau_0}^{\tau_1} d\tau L_{\rm A} 
\label{4.2}
\end{align}
remains invariant under the reparametrization $\tau \rightarrow \tau^{\prime}$. 
Just like $L_{\rm D}$, the Lagrangian $L_{\rm A}$ remains invariant under the gauge transformation 
(\ref{3.4}) and possesses the antisymmetric property 
\begin{align}
L_{\rm A} \left(q_{1}^{\mu}, \dot{q}_{1}^{\mu}, \lambda_{1\mu}; 
q_{2}^{\mu}, \dot{q}_{2}^{\mu}, \lambda_{2\mu} \right)
=-L_{\rm A} \left(q_{2}^{\mu}, \dot{q}_{2}^{\mu}, \lambda_{2\mu}; 
q_{1}^{\mu}, \dot{q}_{1}^{\mu}, \lambda_{1\mu} \right) . 
\label{4.3}
\end{align}

We now derive the Euler-Lagrange equations for the dynamical variables from $L_{\rm A}$.  
The Euler-Lagrange equation for $x^{\mu}$ is found to be 
\begin{align}
\frac{d}{d\tau} \left[ \frac{1}{\left( q_{1}^{2} q_{2}^{2} \right)^{1/4}}
\left(\lambda_{1\mu}-\lambda_{2\mu} \right) \right] 
+\frac{3}{2e \left( q_{1}^{2} q_{2}^{2} \right)^{1/4}} 
\;\! \partial_{\mu} F_{\nu\rho}(x) q_{1}^{\nu} q_{2}^{\rho} =0\,.
\label{4.4}
\end{align}
This equation includes the gauge-invariant combination 
$\lambda_{1\mu}-\lambda_{2\mu}$ owing to the gauge invariance of $L_{\rm A}$,   
and hence cannot determine $\lambda_{1\mu}$ and $\lambda_{2\mu}$ uniquely.

The Euler-Lagrange equation for $q_{1}^{\mu}$ can be written as 
\begin{align}
&\frac{1}{\left( q_{1}^{2} q_{2}^{2} \right)^{1/4}} 
\left[\;\! \frac{1}{2} \left( \frac{d}{d\tau} \frac{\partial K_{1}}{\partial\dot{q}_{1}^{\mu}}
-\frac{\partial K_{1}}{\partial q_{1}^{\mu}} \right) 
-\frac{k}{2} \left( \frac{d}{d\tau} \frac{\partial J}{\partial\dot{q}_{1}^{\mu}}
-\frac{\partial J}{\partial q_{1}^{\mu}} \right)
+\lambda_{1\mu} -\xi_{\mu} +\frac{3}{2e} F_{\mu\nu}(x) q_{2}^{\nu} \:\! \right] 
\notag
\\
&+\left(\frac{d}{d\tau} \frac{1}{\left(q_{1}^{2} q_{2}^{2} \right)^{1/4}} \right) 
\left( \:\! \frac{1}{2} \frac{\partial K_{1}}{\partial\dot{q}_{1}^{\mu}} 
-\frac{k}{2} \frac{\partial J}{\partial\dot{q}_{1}^{\mu}} \right)
+\frac{q_{1\mu}}{2q_{1}^{2}} L_{\rm A} =0 \,,  
\label{4.5}
\end{align}
where $K_{1}$ and $J$ are given by Eq. (\ref{3.8}) and 
\begin{align}
J:=\frac{\dot{q}_{1\perp\mu} q_{2}^{\mu}}{\sqrt{q_{1}^{2}}} 
-\frac{\dot{q}_{2\perp\mu} q_{1}^{\mu}}{\sqrt{q_{2}^{2}}} \,. 
\label{4.6}
\end{align}
Applying the formulas (\ref{3.9})--(\ref{3.11}) and 
\begin{align}
& \frac{\partial J}{\partial \dot{q}_{1}^{\mu}} 
=\frac{1}{\sqrt{q_{1}^{2}}} \left( q_{2\mu}-\frac{q_1 q_2}{q_{1}^{2}} q_{1\mu} \right) ,  
\label{4.7}
\\
&\frac{d}{d\tau} \frac{\partial J}{\partial\dot{q}_{1}^{\mu}}
-\frac{\partial J}{\partial q_{1}^{\mu}} 
=\frac{d}{d\tau} \frac{q_{2\mu}}{\sqrt{q_{2}^{2}}}
+\frac{1}{\sqrt{q_{1}^{2}}} \left( \dot{q}_{2\mu}-\frac{q_1 \dot{q}_2}{q_{1}^{2}} q_{1\mu} \right) 
\label{4.8}
\end{align}
to Eq. (\ref{4.5}) appropriately, we obtain 
\begin{align}
& \frac{k}{2} \left[\:\! \frac{d}{d\tau} \frac{q_{2\mu}}{\sqrt{q_{2}^{2}}} 
+\frac{1}{\sqrt{q_{1}^{2}}} \left( \dot{q}_{2\mu}-\frac{q_1 \dot{q}_2}{q_{1}^{2}} q_{1\mu} \right) \right]
\notag
\\
& =\frac{3}{2e} F_{\mu\nu}(x) q_{2}^{\nu} 
+\frac{\left( q_{1}^{2} q_{2}^{2} \right)^{1/4} q_{1\mu}}{2q_1^2} L_{\rm A} 
+\frac{\ddot{q}_{1\perp\mu}}{q_{1}^{2}} 
-\left( \frac{5q_{1} \dot{q}_{1}}{q_{1}^{2}} +\frac{q_{2} \dot{q}_{2}}{q_{2}^{2}} \right) 
\frac{\dot{q}_{1\perp \mu}}{2q_1^2}
+\lambda_{1\mu} -\xi_{\mu} 
\notag
\\
&\quad \, +\frac{k}{4\sqrt{q_{1}^{2}}}
\left( \frac{q_{1} \dot{q}_{1}}{q_{1}^{2}} +\frac{q_{2} \dot{q}_{2}}{q_{2}^{2}} \right) 
\left( q_{2\mu}-\frac{q_1 q_2}{q_{1}^{2}} q_{1\mu} \right) ,
\label{4.9}
\end{align}
where $q_{1} q_{2} := q_{1\mu} q_{2}^{\mu}$ and $q_{1} \dot{q}_{2} := q_{1\mu} \dot{q}_{2}^{\mu}$. 
Similarly, the Euler-Lagrange equation for $q_{2}^{\mu}$ is derived as  
\begin{align}
& \frac{k}{2} \left[\:\! \frac{d}{d\tau} \frac{q_{1\mu}}{\sqrt{q_{1}^{2}}} 
+\frac{1}{\sqrt{q_{2}^{2}}} \left( \dot{q}_{1\mu}-\frac{\dot{q}_1 q_2}{q_{2}^{2}} q_{2\mu} \right) \right]
\notag
\\
& =\frac{3}{2e} F_{\mu\nu}(x) q_{1}^{\nu} 
-\frac{\left( q_{1}^{2} q_{2}^{2} \right)^{1/4} q_{2\mu}}{2q_2^2} L_{\rm A} 
+\frac{\ddot{q}_{2\perp\mu}}{q_{2}^{2}} 
-\left( \frac{q_{1} \dot{q}_{1}}{q_{1}^{2}} +\frac{5q_{2} \dot{q}_{2}}{q_{2}^{2}} \right) 
\frac{\dot{q}_{2\perp \mu}}{2q_2^2}
+\lambda_{2\mu} -\xi_{\mu}
\notag
\\
&\quad \, +\frac{k}{4\sqrt{q_{2}^{2}}}
\left( \frac{q_{1} \dot{q}_{1}}{q_{1}^{2}} +\frac{q_{2} \dot{q}_{2}}{q_{2}^{2}} \right) 
\left( q_{1\mu}-\frac{q_1 q_2}{q_{2}^{2}} q_{2\mu} \right) . 
\label{4.10}
\end{align}
The Euler-Lagrange equations for $\lambda_{1\mu}$, $\lambda_{2\mu}$, and $\xi_{\mu}$ 
are respectively found to be  
\begin{align}
q_{1}^{\mu}&=\dot{x}^{\mu} ,
\label{4.11}
\\
q_{2}^{\mu}&=\dot{x}^{\mu} , 
\label{4.12}
\\
q_{1}^{\mu}&=q_{2}^{\mu} ,
\label{4.13}
\end{align}
which are compatible with one another.

Substituting Eqs. (\ref{4.11}) and (\ref{4.12}) into Eq. (\ref{4.9}) and noting 
\begin{align}
L_{\rm A} \left(q_{1}^{\mu}, \dot{q}_{1}^{\mu}, \lambda_{1\mu}; 
q_{2}^{\mu}, \dot{q}_{2}^{\mu}, \lambda_{2\mu} \right)
=L_{\rm A} \left(\dot{x}^{\mu}, \ddot{x}^{\mu}, \lambda_{1\mu}; 
\dot{x}^{\mu}, \ddot{x}^{\mu}, \lambda_{2\mu} \right) 
=0\,, 
\label{4.14}
\end{align}
we have 
\begin{align}
k \frac{d}{d\tau} \frac{\dot{x}^{\mu}}{\sqrt{\dot{x}^{2}}} 
&=\frac{3}{2e} F^{\mu\nu}(x) \dot{x}_{\nu}  
+\frac{\dddot{x}_{\perp}^{\mu}}{\dot{x}^{2}} 
-\frac{3 (\dot{x} \ddot{x}) \ddot{x}_{\perp}^{\mu}}{(\dot{x}^{2})^{2}} 
+\lambda_{1}^{\mu} -\xi^{\mu} \,. 
\label{4.15}
\end{align}
Similarly, substituting Eqs. (\ref{4.11}) and (\ref{4.12}) into Eq. (\ref{4.10}) and using (\ref{4.14}), 
we have 
\begin{align}
k \frac{d}{d\tau} \frac{\dot{x}^{\mu}}{\sqrt{\dot{x}^{2}}} 
&=\frac{3}{2e} F^{\mu\nu}(x) \dot{x}_{\nu}  
+\frac{\dddot{x}_{\perp}^{\mu}}{\dot{x}^{2}} 
-\frac{3 (\dot{x} \ddot{x}) \ddot{x}_{\perp}^{\mu}}{(\dot{x}^{2})^{2}} 
+\lambda_{2}^{\mu} -\xi^{\mu} \,. 
\label{4.16}
\end{align}
Comparing Eq. (\ref{4.15}) and Eq. (\ref{4.16}) leads to 
\begin{align}
\lambda_{1}^{\mu}=\lambda_{2}^{\mu} \,.
\label{4.17}
\end{align}
Then we see that Eq. (\ref{4.4}) is identically satisfied owing to  
$\partial_{\mu} F_{\nu\rho}(x) q_{1}^{\nu} q_{2}^{\rho}
=\partial_{\mu} F_{\nu\rho}(x) \dot{x}^{\nu} \dot{x}^{\rho}=0$. 
With $\varLambda^{\mu}:=\lambda^{\mu}-\xi^{\mu}$  
$(^{\:\!}\lambda^{\mu}:=\lambda_{1}^{\mu}=\lambda_{2}^{\mu\;\!})$, 
Eqs. (\ref{4.15}) and (\ref{4.16}) can be written together as a single equation 
\begin{align}
m \frac{d}{d\tau} \frac{\dot{x}^{\mu}}{\sqrt{\dot{x}^{2}}} 
&=eF^{\mu\nu}(x) \dot{x}_{\nu}  
+\frac{2}{3}e^{2} \left( \frac{\dddot{x}_{\perp}^{\mu}}{\dot{x}^{2}} 
-\frac{3 (\dot{x} \ddot{x}) \ddot{x}_{\perp}^{\mu}}{(\dot{x}^{2})^{2}} \right) 
+\frac{2}{3}e^{2} \varLambda^{\mu} ,  
\label{4.18}
\end{align}
after the substitution of $k=3m/2e^{2}$. 
This equation is completely the same as Eq. (\ref{3.23}). 
In this way, it is established that the Lagrangian $L_{\rm A}$ also yields 
the Lorentz-Dirac equation with a source-like term.

\section{Summary and discussion}

We have presented two relativistic Lagrangians $L_{\rm D}$ and $L_{\rm A}$ and have demonstrated   
that the Euler-Lagrange equations derived from $L_{\rm D}$, or those derived from $L_{\rm A}$, 
together lead to the Lorentz-Dirac equation with a source-like term.  
This equation is a differential equation for $x^{\mu}$ having the inhomogeneous 
term $2e^{2} \varLambda^{\mu} /3$. 
Hence it follows that its solutions naturally depend on $\varLambda^{\mu}$. 
The Lorentz-Dirac equation itself can be obtained in a particular situation such that $\varLambda^{\mu}=0$.  
For this reason, $L_{\rm D}$ and $L_{\rm A}$ can simply be said to be the Lagrangians for 
the Lorentz-Dirac equation.

Contracting both sides of Eq. (\ref{4.18}) with $\dot{x}^{\mu}$, we have 
the orthogonality condition 
\begin{align}
\dot{x}_{\mu} \varLambda^{\mu}=0 \,.  
\label{5.1}
\end{align}
This condition can be written as $\varLambda^{0}=\vc{v}\cdot\vc{\varLambda}\s$,  
with $\vc{\varLambda}:=(\varLambda^{r})$ (${}^{\:\!}r=1,2,3 {}^{\:\!}$) and the velocity vector 
$\vc{v}:=( dx^{r}/dx^{0} )^{\s}$. 
Accordingly, the 4-vector $(\varLambda^{\mu})$ can be expressed as $(\vc{v}\cdot\vc{\varLambda}, \vc{\varLambda})$. 
As can be seen from Eq. (\ref{4.18}), the source-like term $2e^{2} \varLambda^{\mu}/3$ is regarded as 
a component of the force 4-vector $(f^{\mu})=(\vc{v}\cdot\vc{f}, \vc{f})$, 
provided that $\vc{f}:=(2e^{2}/3) \vc{\varLambda}$ is identified with an external (non-Lorentzian) force acting on 
the charged particle.  
In this way, the source-like term can be treated as a component of 
the force 4-vector of an external force.

In the indirect formulation of the damped harmonic oscillator \cite{Bateman}, 
a pair of two coordinate variables 
is introduced to describe the motion forward in time and that backward in time. 
A pair of $q_{1}^{\mu}$ and $q_{2}^{\mu}$ does not correspond to such a pair of coordinate variables . 
In fact, $q_{1}^{\mu}$ and $q_{2}^{\mu}$ are included even in the Lagrangian  
with an exponential damping function $L_{\rm D}$.  
Also, $q_{1}^{\mu}=q_{2}^{\mu}=\dot{x}^{\mu}$ is eventually found from the Lagrangians  
$L_{\rm D}$ and $L_{\rm A}$. For these reasons,  
$q_{1}^{\mu}$ and $q_{2}^{\mu}$ should simply be regarded as auxiliary variables useful for 
deriving the Lorentz-Dirac equation.

As has been emphasized above, 
$L_{\rm D}$ explicitly depends on the parameter $\tau$, whereas 
$L_{\rm A}$ does not depend explicitly on $\tau$. 
In a consistent quantization of the damped harmonic oscillator \cite{Dekker, FT, Banerjee}, 
the indirect formulation based on an explicitly time-independent Lagrangian 
is adopted rather than the direct formulation based on an explicitly time-dependent Lagrangian. 
Referring to this fact,  
we should choose $L_{\rm A}$ as a desirable Lagrangian when we consider quantum theory 
of a charged particle described by the Lorentz-Dirac equation. 
The Lagrangian and Hamiltonian formulations based on $L_{\rm A}$ and the subsequent 
quantization procedure are interesting issues that should be addressed in the future.

\section*{Acknowledgments}

We are grateful to Shigefumi Naka, Takeshi Nihei and Akitsugu Miwa   
for their useful comments.  
We thank Keita Seto for valuable information on the Lorentz-Dirac equation. 
One of us (T.S.) thanks Kenji Yamada, Katsuhito Yamaguchi and Haruki Toyoda 
for their encouragement. 
The work of S.D. is supported in part by  
Grant-in-Aid for Fundamental Scientific Research from   
College of Science and Technology, Nihon University.

\appendix
\section{Lorentz-Dirac equation written in terms of an arbitrary world-line parameter}

Let us recall the Lorentz-Dirac equation Eq. (\ref{2}), i.e.,  
\begin{align}
m \frac{du^{\mu}}{dl} =eF^{\mu\nu}(x) {u}_{\nu} 
+\frac{2}{3} e^{2}\big( \delta^{\mu}{}_{\nu} -u^{\mu} u_{\nu} \big) \frac{d^2 u^{\nu}}{dl^{2}} \,,
\label{A1}
\end{align}
which is written in terms of the proper time $l$. 
Noting that the infinitesimal proper time can be expressed as 
$dl=\sqrt{\dot{x}^{2}}\:\!d\tau$, we can show that 
\begin{align}
u^{\mu}:\!&=\frac{dx^{\mu}}{dl}=\frac{\dot{x}^{\mu}}{\sqrt{\dot{x}^{2}}} \,, 
\label{A2}
\\
\frac{du^{\mu}}{dl} &=\frac{\ddot{x}^{\mu}_{\perp}}{\dot{x}^2} \,,
\label{A3}
\\
\frac{d^2 u^{\mu}}{dl^2}
&=\frac{1}{\sqrt{\dot{x}^{2}}} 
\left[\;\! \frac{\dddot{x}^{\mu}_{\perp}}{\dot{x}^{2}}  
-\frac{3(\dot{x}\ddot{x}) \ddot{x}^{\mu}_{\perp}}{(\dot{x}^{2})^{2}} 
-\left(\ddot{x}^{2}-\frac{(\dot{x}\ddot{x})^{2}}{\dot{x}^{2}} \right) 
\frac{\dot{x}^{\mu}}{(\dot{x}^2)^{2}} \right] , 
\label{A4}
\end{align}
with 
\begin{align}
\ddot{x}^{\mu}_{\perp}:=\ddot{x}^{\mu}-\frac{\dot{x}\ddot{x}}{\dot{x}^{2}} \dot{x}^{\mu} , 
\qquad 
\dddot{x}^{\mu}_{\perp}:=\dddot{x}^{\mu}-\frac{\dot{x}\dddot{x}}{\dot{x}^{2}} \dot{x}^{\mu} . 
\label{A5}
\end{align}
Here,  
\begin{align}
\dot{x}^{\mu}:=\frac{dx^{\mu}}{d\tau} \,, 
\qquad 
\ddot{x}^{\mu}:=\frac{d^2 x^{\mu}}{d\tau^2} \,, 
\qquad 
\dddot{x}^{\mu}:=\frac{d^3 x^{\mu}}{d\tau^3} &\,, 
\label{A6}
\\
\dot{x}^{2}:=\dot{x}_{\mu} \dot{x}^{\mu}, 
\qquad 
\ddot{x}^{2}:=\ddot{x}_{\mu} \ddot{x}^{\mu}, 
\qquad 
\dot{x}\ddot{x}:=\dot{x}_{\mu} \ddot{x}^{\mu}, 
\qquad 
\dot{x}\dddot{x} &:=\dot{x}_{\mu} \dddot{x}{}^{\mu}. 
\label{A7}
\end{align}

Substituting Eqs. (\ref{A2}) and (\ref{A4}) into Eq. (\ref{A1}), we obtain
\begin{align}
m \frac{d}{d\tau} \frac{\dot{x}^{\mu}}{\sqrt{\dot{x}^{2}}} 
&=eF^{\mu\nu}(x) \dot{x}_{\nu}  
+\frac{2}{3}e^{2} \left( \frac{\dddot{x}_{\perp}^{\mu}}{\dot{x}^{2}} 
-\frac{3 (\dot{x} \ddot{x}) \ddot{x}_{\perp}^{\mu}}{(\dot{x}^{2})^{2}} \right) .
\label{A8}
\end{align}
This is the Lorentz-Dirac equation written in terms of an arbitrary world-line parameter $\tau$. 
We can directly derive Eq. (\ref{A8}) by evaluating the reaction force 
due to the particle's own electromagnetic radiation 
without adopting the proper-time gauge $\tau=l$.

\bibliographystyle{elsarticle-num}
\bibliography{<your-bib-database>}

\begin{thebibliography}{00}


\bibitem{LL}
L.~D.~Landau and E.~M.~Lifshitz,  
{\it The Classical Theory of Fields}, 4th ed., Butterworth-Heinemann, Oxford, 1975. 

\bibitem{Jackson}
J.~D.~Jackson, 
{\it Classical Electrodynamics}, 3rd ed., John Wiley \& Sons, New York, 1998.

\bibitem{Barut1} 
A.~O.~Barut, 
{\it Electrodynamics and Classical Theory of Fields and Particles}, 
Dover, New York, 1980. 

\bibitem{Rohrlich1}
F.~Rohrlich, 
{\it Classical Charged Particles}, 3rd ed., World Scientific, Singapore, 2007. 

\bibitem{Rohrlich2}
F.~Rohrlich, 
^^ ^^ The dynamics of a charged sphere and the electron," 
Am. J. Phys. {\bf 65} (1997) 1051-1056. 

\bibitem{Spohn1}
H.~Spohn, 
{\it Dynamics of Charged Particles and Their Radiation Field}, 
Cambridge University Press, Cambridge, 2004. 

\bibitem{Lorentz1}
H.~A.~Lorentz,  
^^ ^^ La th\'{e}orie \'{e}lecromagnetique de Maxwell et son application 
aux corps mouvemants," Arch. N\'{e}erland. Sci. Exactes Nat. {\bf 25} (1892) 363-552. 

\bibitem{Abraham}
M.~Abraham, 
{\it Theorie der Elektrizit\"{a}t,} Vol. II: {\it Elektromagnetische Theorie der Strahlung}, 
Teubner, Leipzig, 1905. 

\bibitem{Lorentz2}
H.~A.~Lorentz,  
{\it The Theory of Electrons and Its Applications to the Phenomena of Light and Radiant Heat,}  
2nd ed., Dover, New York, 1952. 

\bibitem{Dirac} 
P.~A.~M.~Dirac, 
^^ ^^ Classical theory of radiating electrons," 
Proc. R. Soc. London Ser. A {\bf 167} (1938) 148-169. 

\bibitem{Barut2} 
A.~O.~Barut, 
^^ ^^ Electrodynamics in terms of retarded fields," 
Phys. Rev. D {\bf 10} (1974) 3335-3336. 

\bibitem{Poisson} 
E.~Poisson, 
^^ ^^ An introduction to the Lorentz-Dirac equation," 
unpublished, arXiv:gr-qc/9912045. 

\bibitem{Nakhleh}
C. W. Nakhleh, 
^^ ^^ The Lorentz-Dirac and Landau-Lifshitz equations  
from the perspective of modern renormalization theory," 
Am. J. Phys. {\bf 81} (2013) 180-185, arXiv:1207.1745 [physics.class-ph]. 

\bibitem{Yaghjian}
A.~D.~Yaghjian, 
{\it Relativistic Dynamics of a Charged Sphere}, 2nd ed., Lect. Notes Phys. Vol. 686, Springer-Verlag, New York, 2006.

\bibitem{Barut3} 
A.~O.~Barut, 
^^ ^^ Renormalization and elimination of preacceleration and runaway solutions of 
the Lorentz-Dirac equation," 
Phys. Lett. A {\bf 145} (1990) 387-390. 

\bibitem{FOC1}
G.~W.~Ford and R. F. O'Connell, 
^^ ^^ Radiation reaction in electrodynamics and the elimination of runaway solutions," 
Phys. Lett. A {\bf 157} (1991) 217-220. 

\bibitem{FOC2}
G.~W.~Ford and R. F. O'Connell, 
^^ ^^ Relativistic form of radiation reaction," 
Phys. Lett. A {\bf 174} (1993) 182-184. 

\bibitem{Spohn2}
H. Spohn, 
^^ ^^ The critical manifold of the Lorentz-Dirac equation," 
Europhys. Lett. {\bf 50} (2000) 287-292, arXiv:physics/9911027. 

\bibitem{Rohrlich3}
F.~Rohrlich, 
^^ ^^ The correct equation of motion of a classical point charge," 
Phys. Lett. A {\bf 283} (2001) 276-278. 

\bibitem{Rohrlich4}
F.~Rohrlich, 
^^ ^^ Dynamics of a charged particle,"  
Phys. Rev. E {\bf 77} (2008) 046609 (1-4). 

\bibitem{Seto}
K. Seto, S. Zhang, J. Koga, H. Nagatomo, M. Nakai, and K. Mima, 
^^ ^^ Stabilization of radiation reaction with vacuum polarization," 
Prog. Theor. Exp. Phys. {\bf 2014}  (2014) 043A01 (1-10), arXiv:1310.6646 [physics.plasm-ph]. 

\bibitem{Carati} 
A.~Carati, 
^^ ^^ A Lagrangian formulation for the Abraham-Lorentz-Dirac equation," 
in {\em Symmetry and Perturbation Theory}  
(Proceedings edited by D. Bambusi and G. Gaeta) 
Consiglio Nazionale delle Ricerche as ^^ ^^ Quaderno GNFM-CNR, Vol. 54", Roma, 1998. 

\bibitem{BM} 
P. M. V. B. Barone and A. C. R. Mendes, 
^^ ^^ Lagrangian description of the radiation damping," 
Phys. Lett. A {\bf 364} (2007) 438-440, arXiv:cond-mat/0607370.  

\bibitem{Kupriyanov} 
V. G. Kupriyanov, 
^^ ^^ Hamiltonian formulation and action principle for the Lorentz-Dirac system," 
Int. J. Theor. Phys. {\bf 45} (2006) 1129-1144. 

\bibitem{Dekker}
H. Dekker, 
^^ ^^ Classical and quantum mechanics of the damped harmonic oscillator," 
Phys. Rep. {\bf 80} (1981) 1-112. 

\bibitem{Caldirola}
P. Caldirola, 
^^ ^^ Forze non conservative nella meccanica quantistica,"
Nuovo Cimento {\bf 18} (1941) 393-400. 

\bibitem{Kanai}
E. Kanai, 
^^ ^^ On the quantization of the dissipative systems," 
Prog. Theor. Phys. {\bf 3} (1948) 440-442. 

\bibitem{Bateman} 
H. Bateman, 
^^ ^^ On dissipative systems and related variational principles," 
Phys. Rev. {\bf 38} (1931) 815-819. 

\bibitem{FT}
H. Feshbach and Y. Tikochinsky, 
^^ ^^ Quantization of the damped harmonic oscillator," 
Trans. N.Y. Acad. Sci., Ser. II {\bf 38} (1977) 44-53. 

\bibitem{Banerjee} 
R. Banerjee and P. Mukherjee, 
^^ ^^ A canonical approach to the quantization of the damped harmonic oscillator,"  
J. Phys. A: Math. Gen. {\bf 35} (2002) 5591-5598, arXiv:quant-ph/0108055. 

\bibitem{Pav1} 
M.~Pav\v{s}i\v{c},  
^^ ^^ Classical motion of membranes, strings and point particles with extrinsic curvature,"  
Phys. Lett. B {\bf 205} (1988) 231-236. 

\bibitem{Pav2} 
M.~Pav\v{s}i\v{c},  
^^ ^^ The quantization of a point particle with extrinsic curvature leads to the Dirac equation,"  
Phys. Lett. B {\bf 221} (1989) 264-268. 

\bibitem{Plyushchay}
M.~S.~Plyushchay,  
^^ ^^ Does the quantization of a particle with curvature lead to Dirac equation?,"  
Phys. Lett. B {\bf 253} (1991) 50-55. 

\end{thebibliography}

\end{document}